\documentclass{appolb}
\usepackage{graphicx}
\usepackage{cite}
\usepackage{amsmath,amssymb} 

\begin{document}
\title{PHENIX results on collision energy dependent L\'evy HBT correlations from $\sqrt{s_{NN}}$ = 15 to 200 GeV%
\thanks{Presented at the XIII Workshop on Particle Correlations and Femtoscopy, \\22-26 May 2018, Krakow, Poland
}
}
\author{D\'aniel Kincses for the PHENIX Collaboration
\thanks{This research was supported by the \'UNKP-18-3 New National Excellence Program of the hungarian Ministry of Human Capacities, as well as the NKFIH grant FK 123842 and the funding agencies listed in Ref. \cite{funding}.}
\address{E\"otv\"os Lor\'and University, P\'azm\'any P\'eter s\'et\'any 1/A, 1117 Budapest, Hungary}
%\address{kincses@ttk.elte.hu}
}
\pagestyle{plain}
\maketitle
\begin{abstract}

Different regions on the QCD phase diagram can be investigated by varying the collision energy and the centrality in heavy-ion collisions. In our latest measurements at the PHENIX experiment at RHIC, we utilize L\'evy-type sources to describe the measured HBT correlation functions. In this paper we report the current status of the analysis of the centrality and beam energy dependence of the L\'evy source parameters in Au+Au collisions from $\sqrt{s_{NN}}$ = 15 GeV to $\sqrt{s_{NN}}$ = 200 GeV.

\end{abstract}

\section{Introduction}

\begin{figure}[htb]
\centerline{\includegraphics[width=0.45\textwidth]{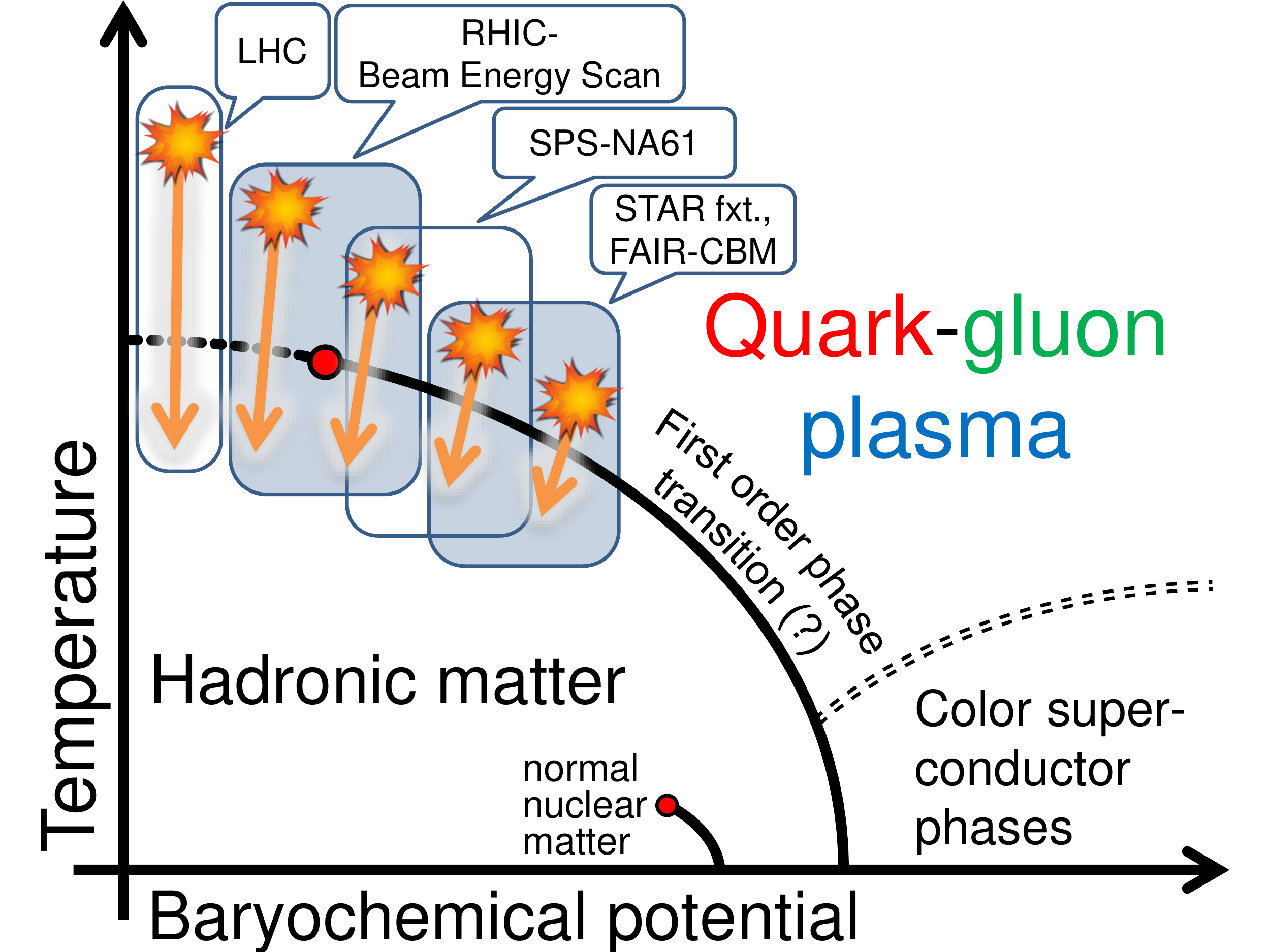}
            \includegraphics[width=0.53\textwidth]{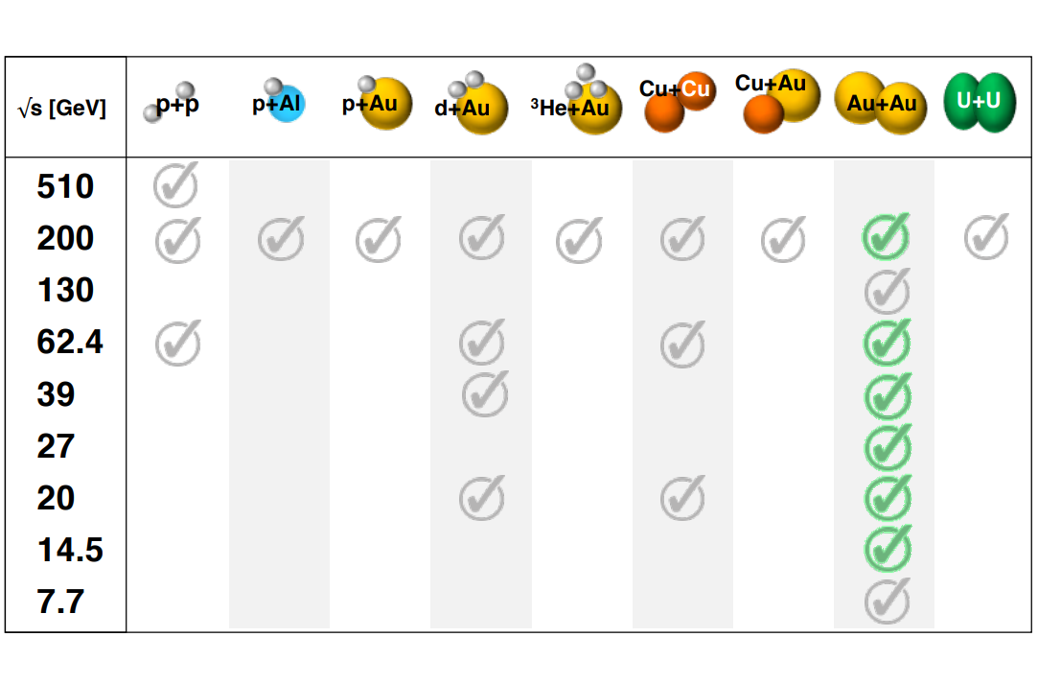}}
\caption{QCD phase diagram and the RHIC Beam Energy Scan}
\label{ff:pd}
\end{figure}

To experimentally explore the phase diagram of strongly interacting matter, one has to investigate heavy-ion collision data from different center of mass collision energies. The aim of the Beam Energy Scan program at the Relativistic Heavy Ion Collider (RHIC) was to collect a huge variety of data from different energies and different kinds of colliding nuclei. The second phase of this program, the BES-II is still ongoing. Complemented with the fixed-target program at STAR the beam energy scan will cover a wide range in baryochemical potential, spanning from about $\mu_B$ = 20 MeV up to $\mu_B$ = 720 MeV \cite{Keane:2017kdq}. For the analysis reported in this paper we used BES-I data recorded by the PHENIX experiment at six different collision energies. Fig. \ref{ff:pd}. illustrates the QCD phase diagram, and the data recorded by PHENIX during BES-I. The datasets used in our analysis are highlighted in the table. 

Having a huge variety of data is necessary but not sufficient for exploring the phase diagram. We also need experimental observables that provide us with information on the nature of the phase transition. One of these useful observables is the quantum-statistical (also called Bose-Einstein or HBT) correlations of identical bosons. With the help of these correlation functions, we can gain information on the space-time structure of the particle-emitting source in heavy-ion collisions. To describe the measured correlation functions one has to utilize a certain type of source function. In the previous years the usual practice was to use a Gaussian-type of source but the latest results from different experiments (PHENIX \cite{Adare:2017vig}, CMS \cite{Sikler:2017mde}) showed that we need to go beyond the Gaussian approximation. Different scenarios (e.g. anomalous diffusion) can lead to the appearance of L\'evy-type sources, as described in Refs. \cite{Csanad:2007fr, Adare:2017vig}. The symmetric L\'evy distribution is the generalization of the Gaussian, and it is characterized by two main parameters, the L\'evy scale $R$, and the L\'evy exponent $\alpha$. It is defined by the following expression:
\begin{equation}
\displaystyle \mathcal{L}(\alpha,R,\boldsymbol{r})=\frac{1}{(2\pi)^3} \int d^3\boldsymbol{q} e^{i\boldsymbol{qr}} e^{-\frac{1}{2}|\boldsymbol{q}R|^{\alpha}}.
\end{equation}
The $R$ parameter is in connection with the physical size of the source, and $\alpha$ parameter determines the shape. If the L\'evy exponent takes the value of 2, we get back the Gaussian distribution, and in case of $\alpha < 2$ the L\'evy distribution exhibits a power-law behavior. If we neglect the final state interactions and assume a spherically symmetric L\'evy source, the correlation function takes the simple form of
\begin{equation}
C(Q) = 1+\lambda\cdot\exp{\left(-(RQ)^\alpha\right)}.
\label{eq:cf}
\end{equation}
%as illustrated on Fig.\ref{ff:cor}. 

%\begin{figure}[htb]
%\centerline{\includegraphics[width=0.49\textwidth]{corrfunc.pdf}}
%\caption{Illustration of a L\'evy-type correlation function}
%\label{ff:cor}
%\end{figure}

In our analysis we defined the one-dimensional relative-momentum variable $Q$ as the absolute value of the three-momentum difference in the LCMS frame, see details in \cite{Adare:2017vig}. In the framework of the Core-Halo model \cite{Bolz:1992hc} the $\lambda$ parameter (the strength of the correlation function) is in connection with the core-halo ratio:
\begin{equation}
\lambda = (N_C/(N_C+N_H))^2.
\end{equation}
In this framework we split the source function into two parts. The core part contains the pions created directly from the freezeout (or from very short lived resonances), while the halo part contains the pions created in decays of long-lived resonances. The number of pions associated with the core and halo part are denoted by $N_C$, and $N_H$ respectively. 

Besides describing data better than the Gaussian, a L\'evy-type source has further importance in the search for the critical point. It can be shown that the L\'evy exponent $\alpha$ parameter is connected to the critical exponent $\eta$ known from statistical physics. In case of a second order phase transition at the critical point, the spatial correlations are characterized by a power-law behavior with an exponent $-(d-2+\eta)$, where $d$ is the dimension of the system. The power-law tail of a three-dimensional L\'evy distribution is described with an exponent of $-1-\alpha$, so just by simply comparing the two cases, we can see that in case of three dimensions (and assuming infinite volume nuclear matter) the exponents are identical at the critical point. If we assume that the universality class of QCD is identical to that of the three-dimensional Ising model (or the random-field 3D Ising model), we can look up the value of the $\eta$ exponent from the literature, and find the rather small value of $\eta = $ 0.03631(3) \cite{El-Showk2014} (or $\eta = 0.5\pm0.05$ in case of the random-field 3D Ising model \cite{PhysRevB.52.6659}). If the previous arguments about the connection between $\eta$ and $\alpha$ are true, with the measurement of $\alpha$ at different center of mass collision energies we may get closer to being able to locate the critical point on the phase diagram of strongly interacting matter.

\section{Centrality dependence of the source parameters in Au+Au collisions at $\sqrt{s_{NN}}$ = 200 GeV}

It has been a matter of discussion for a while now, that finite size and finite time effects play an important role in the phase transitions that occur in heavy ion collisions \cite{Lacey:2015yxg}. It is obvious that we cannot talk about infinite volume nuclear matter in case of such collisions, but we may be able to control the system size (with the help of the centrality of these collisions). For the previously mentioned reasons it is important to investigate in detail the centrality dependence of the extracted source parameters. In this section we present such an analysis in $\sqrt{s_{NN}}$ = 200 GeV Au+Au collisions (recorded by PHENIX during the 2010 running period).

\begin{figure}[htb]
	\centerline{
	\includegraphics[clip,trim=0cm 1.4cm 2.00cm 0.8cm,width=0.49\textwidth]{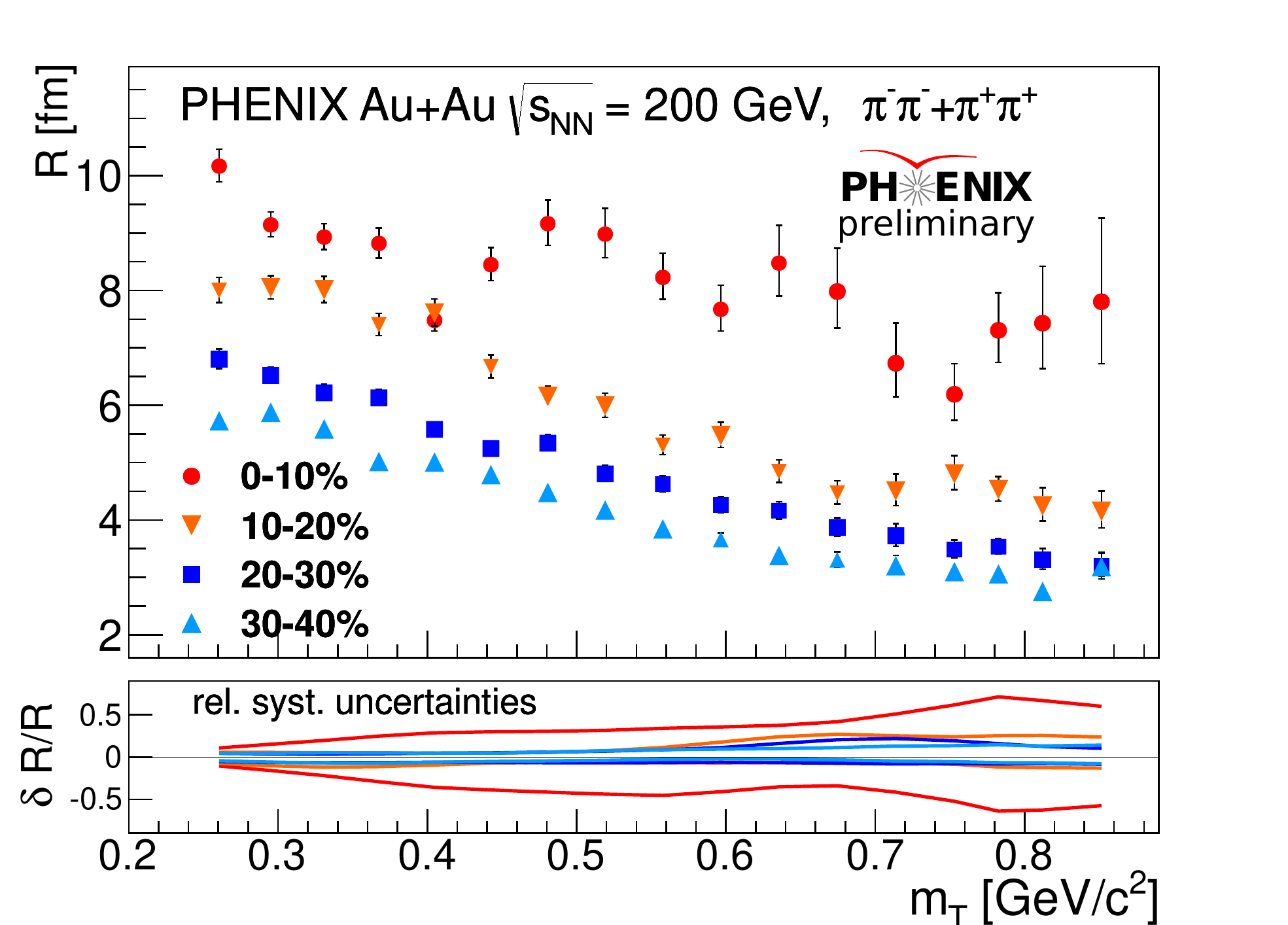}
	\includegraphics[clip,trim=0cm 1.4cm 2.00cm 0.8cm,width=0.49\textwidth]{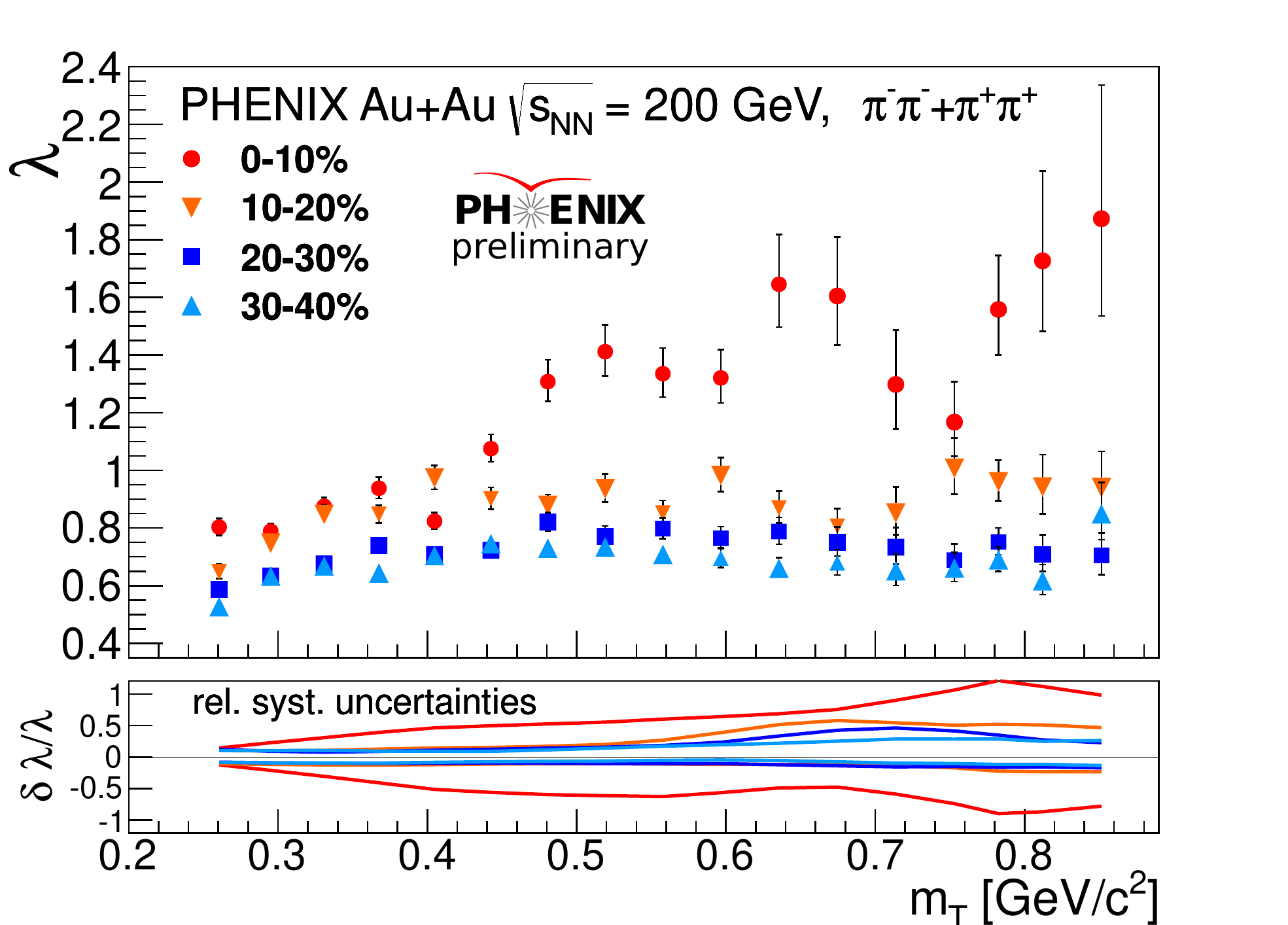}}
	\centerline{
	\includegraphics[clip,trim=0cm 0.cm 2.00cm 0.8cm,width=0.49\textwidth]{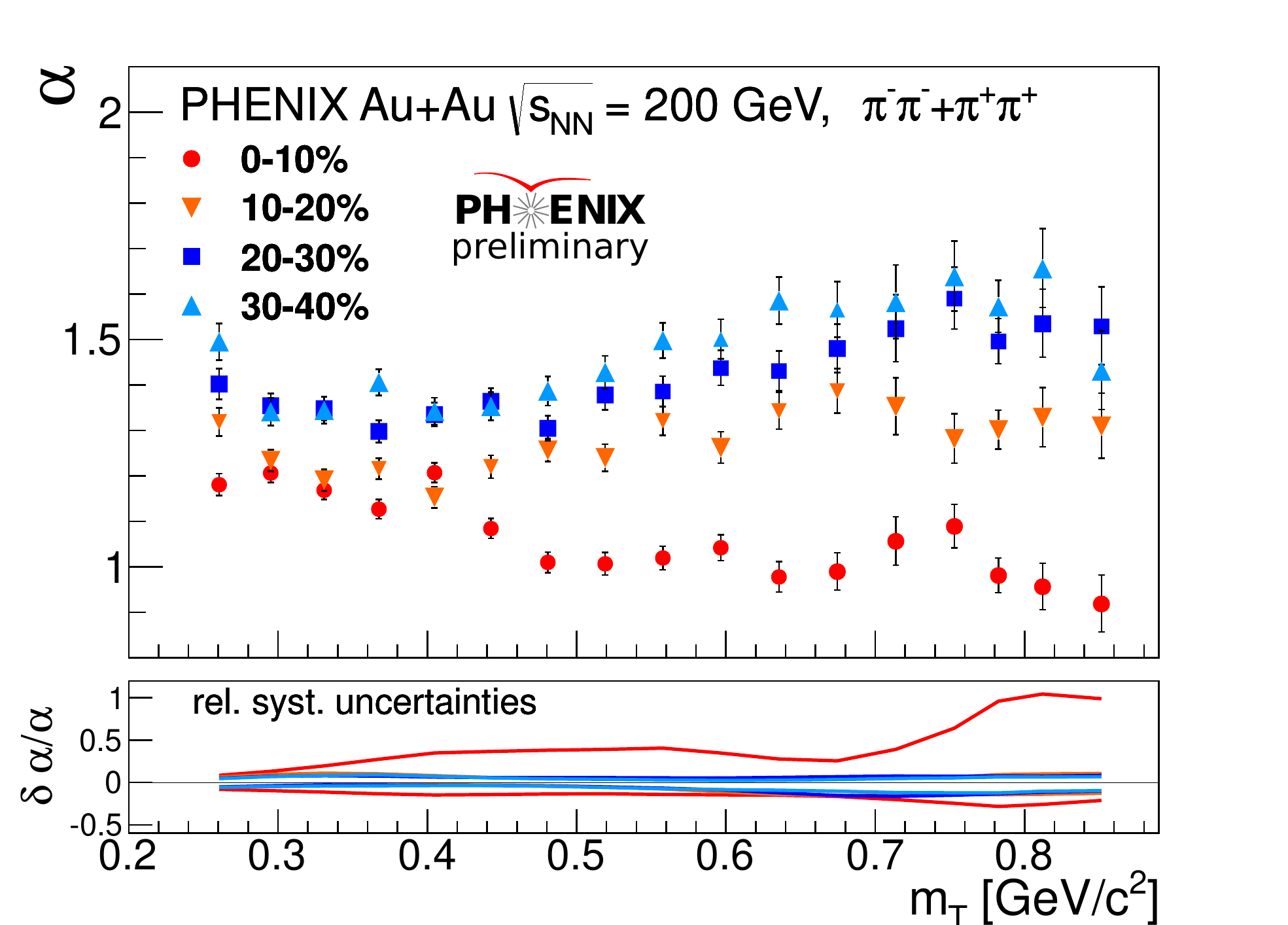}
	\includegraphics[clip,trim=0cm 0.cm 2.00cm 0.8cm,width=0.49\textwidth]{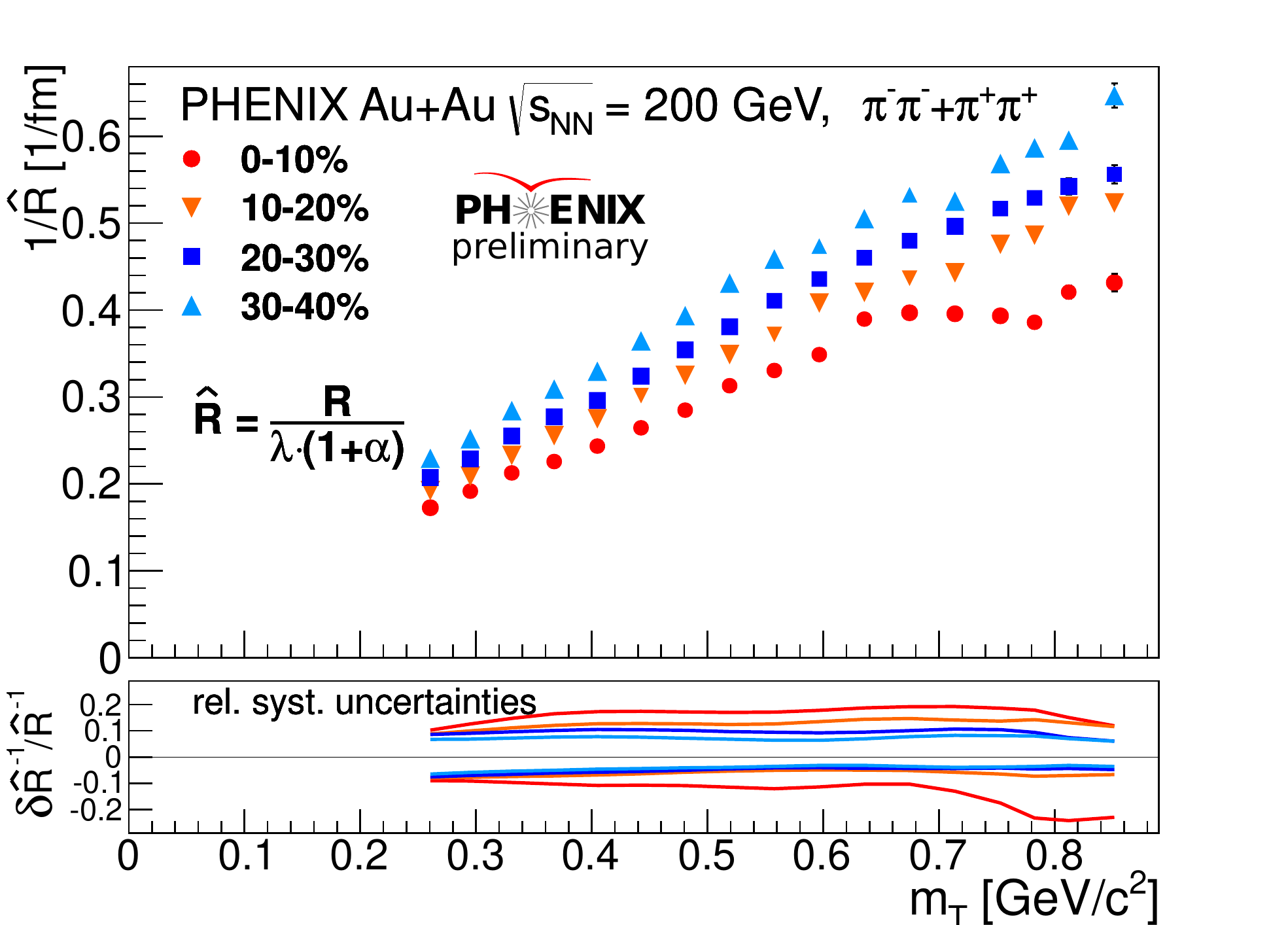}}
	\caption{Centrality and $m_T$ dependence of L\'evy source parameters in $\sqrt{s_{NN}}$ = 200 GeV Au+Au collisions. The different colors and marker styles are denoting the different centrality ranges. The auxiliary figures at the bottom show the relative systematic uncertainties.}
	\label{ff:centdep}
	\end{figure}

In our analysis we measured one-dimensional two-pion HBT correlation functions (for both $\pi^+\pi^+$ and $\pi^-\pi^-$ pairs). After the experimental construction of the correlation functions with the help of the usual event-mixing method, we confirmed that the $\pi^+\pi^+$ and $\pi^-\pi^-$ data are consistent with each other, as expected theoretically. We combined the two in order to increase statistical precision. We used 17 different average transverse mass ($m_T$) bins, and 4 centrality intervals with 10\% increments. The previously shown formula (Eq. (\ref{eq:cf})) does not take into account final state interactions. In the fitting process we incorporated the final-state Coulomb effect in our fit function, and introduced a linear background (which was consistent with a constant in most cases). The extracted fit parameters are shown on Fig.~\ref{ff:centdep}.

The L\'evy scale parameter, $R$, shows the distinctive features of a geometrical centrality dependence -- its value decreases as we go from central to peripheral events. It also shows the usual decrease as we go to higher average transverse mass values, which is consistent with expectations from hydrodynamics. The strength of the correlation function, $\lambda$, decreases at small $m_T$, which could be a possible sign of in-medium mass modification, for details see Refs. \cite{Adare:2017vig,Vance:1998wd}. The L\'evy exponent $\alpha$ parameter is between 0.5 and 2 in all centrality and $m_T$ bins, so it is clearly far from both the Gaussian ($\alpha = 2$) and the conjectured CEP value ($\alpha\lesssim 0.5$ \cite{Csorgo:2005it}). 

During the fitting of the correlation functions we observed that the parameters are strongly correlated, which motivated us to look for less correlated combinations. We indeed found such a combination empirically, defined as $\widehat{R} = R/(\lambda\cdot(1+\alpha))$. This parameter is less correlated with the others, and it can be measured much more precisely. What is quite fascinating is the linear dependence of $1/\widehat{R}$ on $m_T$. The possible physical interpretation of this observation is still unclear. More details on the centrality dependence of the parameters can be found in Refs. \cite{Lokos:2018dqq, Kincses:2017zlb}.

\section{Excitation function of the source parameters}

\begin{figure}[htb]
	\centerline{
	\includegraphics[clip,trim=0cm 1.45cm 2.00cm 0.8cm,width=0.49\textwidth]{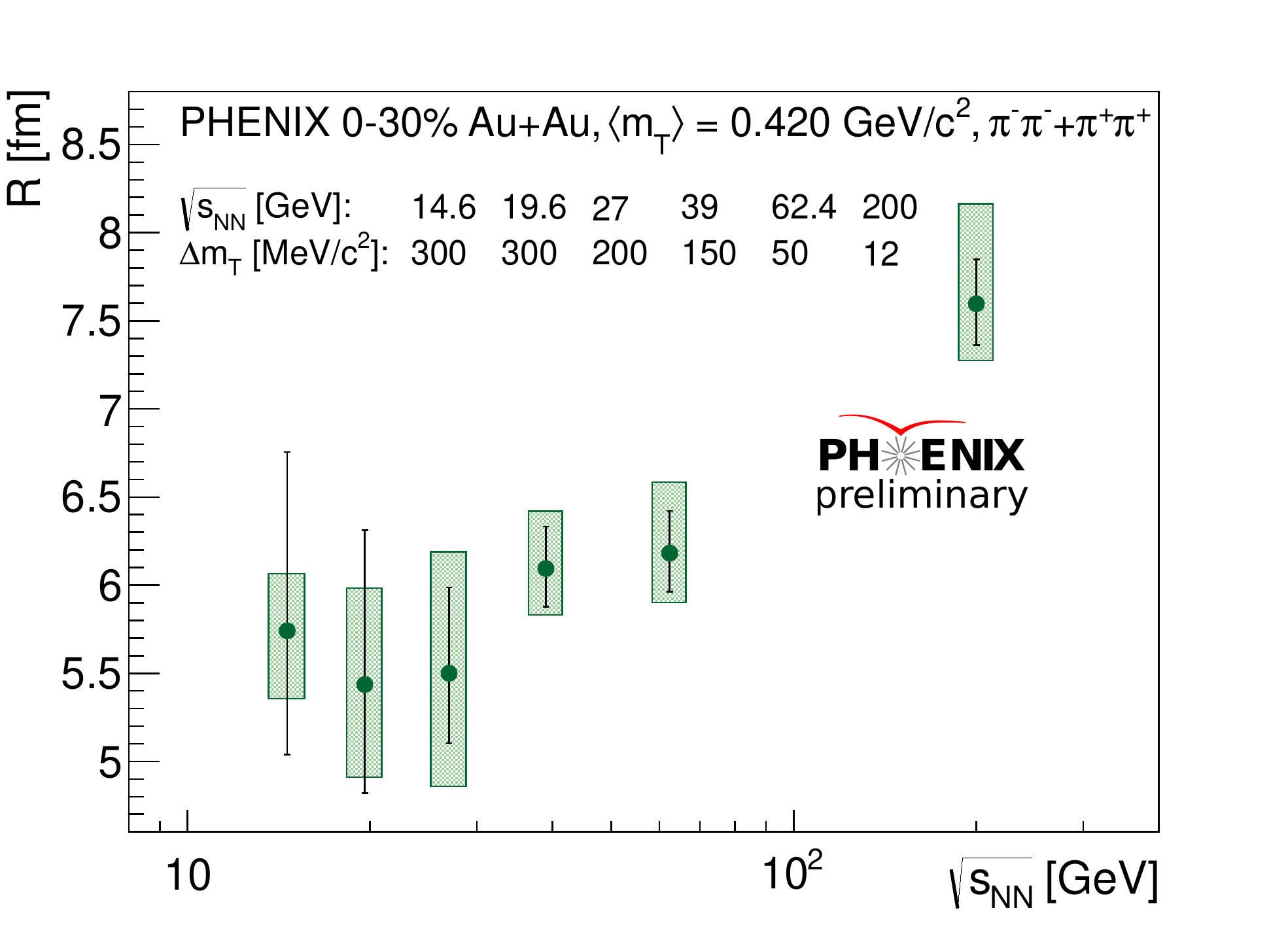}
	\includegraphics[clip,trim=0cm 1.45cm 2.00cm 0.8cm,width=0.49\textwidth]{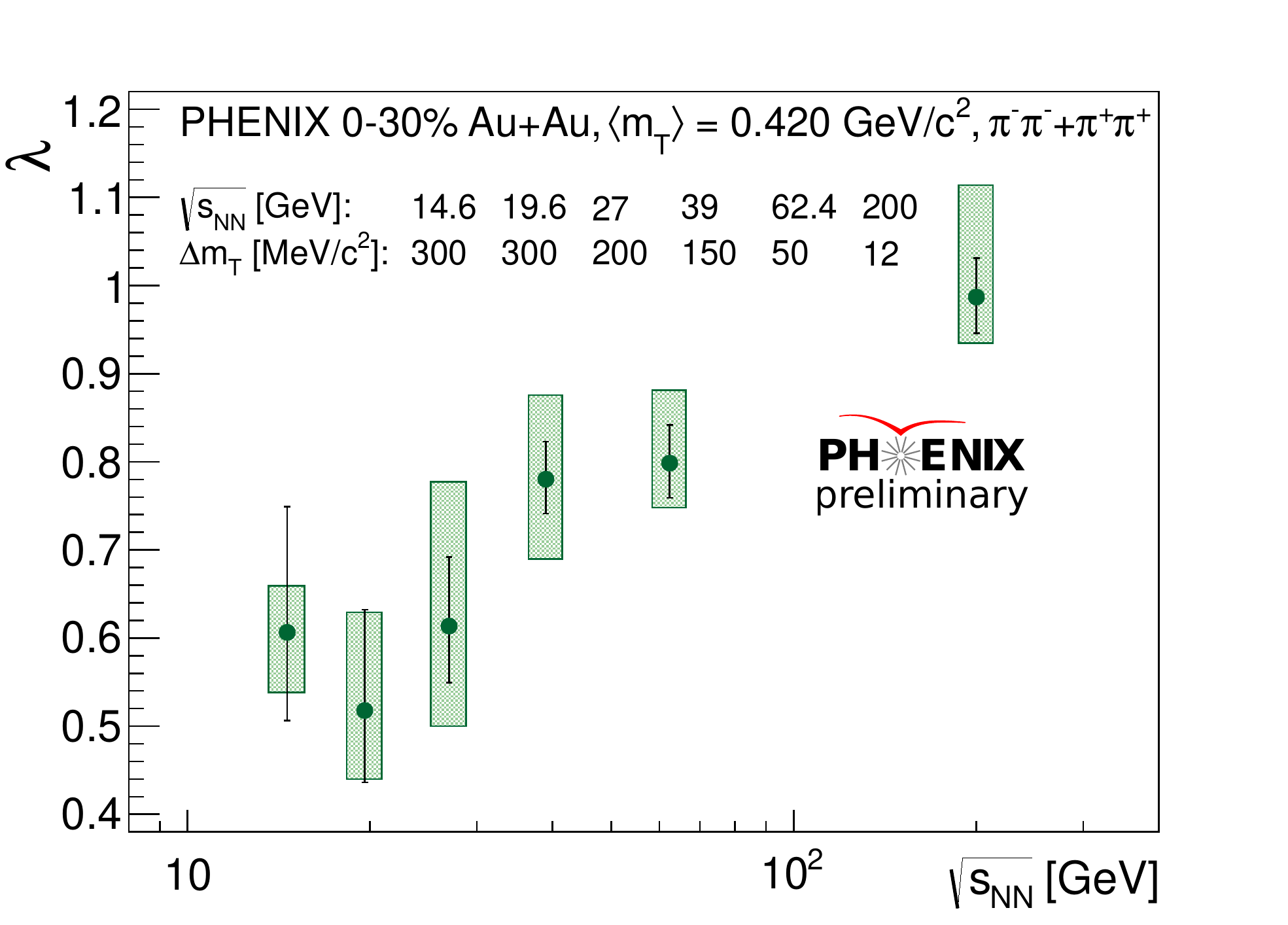}}
	\centerline{
	\includegraphics[clip,trim=0cm 0.cm 2.00cm 0.8cm,width=0.49\textwidth]{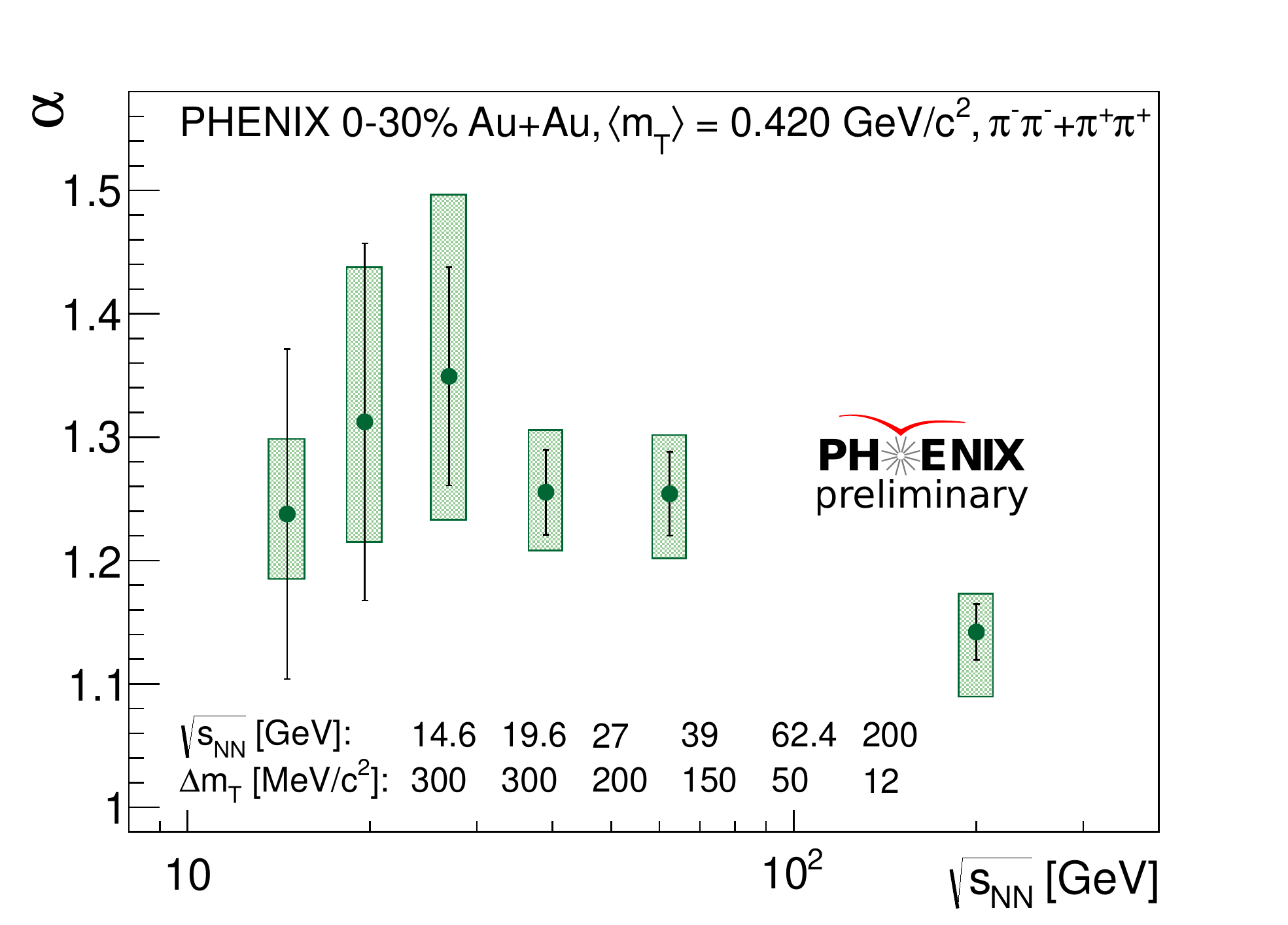}
	\includegraphics[clip,trim=0cm 0.cm 2.00cm 0.8cm,width=0.49\textwidth]{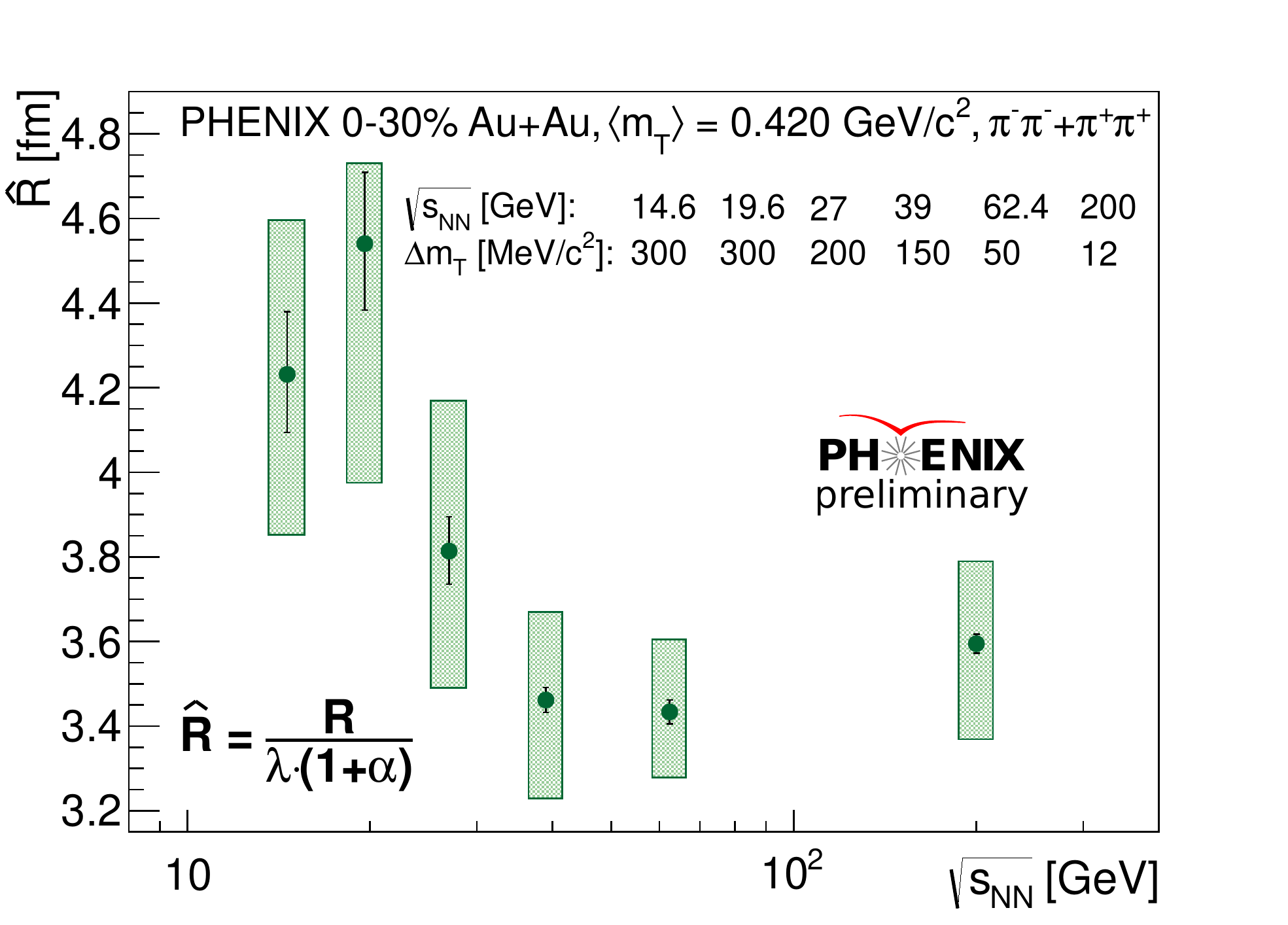}}
	\caption{Excitation functions of L\'evy source parameters in Au+Au collisions. The errorbars represent the statistical uncertainties, while the green boxes show the systematic uncertainties.}
	\label{ff:exc}
	\end{figure}

In case of the $\sqrt{s_{NN}} = 200$ GeV data the investigation of detailed centrality and $m_T$ dependence becomes possible due to the adequate statistical precision. However, when we go down in energy, doing a similar analysis can become quite challenging due to the low statistics. To overcome this issue, we chose to concentrate only on the excitation function of the parameters, and used only one fairly wide centrality range (0-30\%), and just one $m_T$ bin. At lower energies we used wider $m_T$ ranges while keeping the mean value, $\langle m_T \rangle$ the the same. This made it possible to perform our measurements on six different energies, $200, 62, 39, 27, 19$, and 15 GeV. We observed that the choice of the $m_T$ bin width can be a non-negligible source of systematic uncertainty. The results on the excitation functions of the L\'evy source parameters can be seen on Fig. \ref{ff:exc}.

One main observation that we can make is that all the parameters show a weak non-monotonicity with collision energy. It is important to note however, that the statistical and systematic uncertainties become quite large at lower energies. It is also important to note that the L\'evy scale $\alpha$ parameter is above 1 at the whole energy range so we are still far from the conjectured critical point value. Among all parameters, the previously discussed scaling parameter $\widehat{R}$ shows the statistically most significant change with energy. 

	\vspace{-1pt}
\section{Summary and outlook}

In these proceedings we presented L\'evy-type HBT measurements at PHENIX. We discussed the detailed centrality and $m_T$ dependence of the L\'evy source parameters in case of the $\sqrt{s_{NN}} = $ 200 GeV data, as well as the excitation functions of the parameters in one given $m_T$ and centrality range. The shown results are still preliminary, and more detailed investigations are currently ongoing. In the future we also plan to finalize other L\'evy HBT results on three-dimensional as well as three-particle correlation analysis.


\begin{thebibliography}{10}

\bibitem{Keane:2017kdq}
D. Keane, J. Phys. Conf. Ser. {\bf 878}, 012015 (2017).

\bibitem{Adare:2017vig}%PPG194
A. Adare {\it et~al.} [PHENIX Coll.], Phys. Rev. {\bf C97}, 064911 (2018).

\bibitem{Sikler:2017mde}
{F. Sikl\'er for the CMS Coll.}, Universe {\bf 3}, 76 (2017).

\bibitem{Csanad:2007fr}%Levy, anomdiff
{M. Csan\'ad, T. Cs\"org\H{o}, and M. Nagy}, Braz. J. Phys. {\bf 37}, 1002 (2007).

\bibitem{Bolz:1992hc}
{J. Bolz {\it et~al.}}, Phys. Rev. {\bf D47}, 3860 (1993).

\bibitem{El-Showk2014}
{S. El-Showk {\it et~al.}}, J. Stat. Phys. {\bf 157}, 869 (2014).

\bibitem{PhysRevB.52.6659}
{H. Rieger}, Phys. Rev. {\bf B52}, 6659 (1995).

\bibitem{Lacey:2015yxg}
{R. A. Lacey}, Nucl. Phys. {\bf A956}, 348 (2016).

\bibitem{Vance:1998wd}
{S. E. Vance, T. Cs\"org\H{o}, D. Kharzeev}, Phys. Rev. Lett. {\bf 81}, 2205 (1998).

\bibitem{Csorgo:2005it}
{T. Cs\"org\H{o}, S. Hegyi, T. Nov\'ak, W. A. Zajc}, AIP Conf. Proc. {\bf 828}, 525 (2006).

\bibitem{Lokos:2018dqq}
{S. L{\"o}k{\"o}s} for the PHENIX Coll., Universe {\bf 4}, 31 (2018).

\bibitem{Kincses:2017zlb}
{D. Kincses for the PHENIX Coll.}, Universe {\bf 4}, 11 (2018).

\bibitem{funding}
PHENIX Funding Agencies: {\tt http://www.bnl.gov/rhic/PHENIX.asp}.


\end{thebibliography}
\end{document}